\documentclass[runningheads]{llncs}
\usepackage{graphicx}
\usepackage{hyperref}

\usepackage{amsfonts, amsmath, amssymb, bbm, bm}

\usepackage{algorithm}
\usepackage{algpseudocode}
\usepackage{booktabs,caption,subcaption}
\usepackage{multirow}

\DeclareMathOperator*{\argmin}{argmin}
\usepackage{xcolor}
\usepackage{array}
\newcolumntype{P}[1]{>{\centering\arraybackslash}p{#1}}

\begin{document}
\title{OTRE: Where Optimal Transport Guided Unpaired Image-to-Image Translation Meets Regularization by Enhancing}
%
\titlerunning{OTRE: Optimal Transport and Regularization by Enhancing}
%
%
\authorrunning{Wenhui. Zhu et al.}
\author{Wenhui Zhu$^{1*}$, Peijie Qiu$^{2*}$, Oana M. Dumitrascu$^3$, Jacob M. Sobczak$^3$, Mohammad Farazi$^{1}$, Zhangsihao Yang$^{1}$, Keshav Nandakumar$^{1}$, Yalin Wang$^1$}
\institute{$^1$ School of Computing and Augmented Intelligence, Arizona State Univ., AZ, USA \\
$^2$ McKeley School of Engineering, Washington Univ. in St. Louis, St. Louis, MO, USA \\
$^3$ Department of Neurology, Mayo Clinic, Scottsdale, AZ, USA}

%
\maketitle              
\def\thefootnote{*}\footnotetext{The two authors contributed equally to this paper.}
\def\thefootnote{}\footnotetext{Our code is publicly available at \href{https://github.com/Retinal-Research/REDOT}{https://github.com/Retinal-Research/REDOT}.}

\begin{abstract}
Non-mydriatic retinal color fundus photography (CFP) is widely available due to the advantage of not requiring pupillary dilation, however, is prone to poor quality due to operators, systemic imperfections, or patient-related causes. Optimal retinal image quality is mandated for accurate medical diagnoses and automated analyses. Herein, we leveraged the \emph{Optimal Transport (OT)} theory to propose an unpaired image-to-image translation scheme for mapping low-quality retinal CFPs to high-quality counterparts. Furthermore, to improve the flexibility, robustness, and applicability of our image enhancement pipeline in the clinical practice, we generalized a state-of-the-art model-based image reconstruction method, regularization by denoising, by plugging in priors learned by our OT-guided image-to-image translation network. We named it as \emph{regularization by enhancing (RE)}. We validated the integrated framework, OTRE, on three publicly available retinal image datasets by assessing the quality after enhancement and their performance on various downstream tasks, including diabetic retinopathy grading, vessel segmentation, and diabetic lesion segmentation. The experimental results demonstrated the superiority of our proposed framework over some state-of-the-art unsupervised competitors and a state-of-the-art supervised method. 
\end{abstract}
\section{Introduction}
Retinal color fundus photography (CFP) is widely and routinely used to diagnose various ocular diseases. Automated analyses are being developed for point-of-care disease screening, based on non-mydriatic CFP~\cite{OANA3}, Furthermore, research is conducted to unlock the CFPs potential to screen for neurodegenerative disorders such as Alzheimer’s disease~\cite{Oana}. Both human and computer-aided analysis methods prefer operating on high-quality retinal CFPs. Whereas patient and provider-friendly, non-mydriatic retinal CFP is prone to noise, e.g., shading artifacts and blurring because of light transmission disturbance, defocusing, abnormal pupils, or suboptimal human operations~\cite{shen2020modeling}, resulting in low-quality CFPs. CFP degradation such as obscuration of blood vessels, and missing or artifactual new lesions, leads to inaccurate diagnostic interpretation. Enhancing low-quality retinal CFPs into high-quality counterparts is of key importance for many downstream tasks, e.g., diabetic retinopathy (DR), blood vessel segmentation, DR lesion segmentation, etc. Shen et al.~\cite{shen2020modeling} proposed a clinically oriented fundus enhancement network (cofe-Net) by inputting pairs of degraded images synthesized by a fundus degradation model and clean, high-quality images. However, collecting paired noisy-clean retinal training data is difficult and expensive in reality. Lehtinen et al.~\cite{n2n} relaxed the paired noisy-clean images as noisy image pairs obtained from the same condition by arguing that the noisy data approaches the clean data on expectation. Krull et al.~\cite{n2v} extended ~\cite{n2n} to a self-supervised training scheme by predicting surrounding pixels around a blind-spot pixel. This family of self-supervised methods strongly assumed that the ``noise" causing the degradation of images was pixel/image-independent. However, unlike natural images, the composition of noise in retinal fundus images was more complicated, therefore, more challenging to model.
\vspace{-0.1em}

Unsupervised methods have recently attracted much attention. Such a process was usually modeled as an end-to-end image-to-image translation task. Many previous explorations~\cite{UNDOMAIN,pix2pix,cyclegan,couple} in this task were built on top of generative adversarial networks(GANs) to map a source domain $Y$ to a target domain $X$. Such a mapping became more challenging when the input and target were unpaired due to extensive mapping ambiguities. To reduce the large searching space, \cite{couple,UNDOMAIN} regularized the GAN by a task-specific regularization, while the CycleGAN~\cite{cyclegan} proposed a generalized regularization called cycle consistency. However, the expensive computation, destruction of lesions, and introduction of non-existing vessel structures limited the application of CycleGAN to retinal fundus images. To reduce the computational complexity of the CycleGAN, Wang et al.~\cite{9763342} proposed an OT-guided GAN (OTTGAN) for unsupervised image denoising with a single generator and discriminator. Although it achieved very significant results in natural image denoising, its adoption of mean squared error cost as the metric and its lack of high-quality image consistency led to the destruction or over-tampering of the vessel and lesion structures in our tasks.

Here, we proposed an integrated unsupervised end-to-end image enhancement framework based on optimal transport (OT) and regularization by denoising~\cite{red} methods. Our novel OT formulation maximally preserves structural consistency (e.g., lesions, vessel structures, optical discs) between enhanced and low-quality images to prevent over-tampering of important structures. To further improve flexibility and robustness to images from different distributions and applicability in real clinical practice where no sufficient data is available to train the model~\cite{pnp,red,8253590}, we refined the enhanced images by our proposed regularization by enhancing (RE), a variant of regularization by denoising (RED) method~\cite{red}, whose priors were learned by the OT-guided network. The contributions of this paper are three folded: 
(1) We proposed a novel OT-guided GAN-based unsupervised end-to-end retinal image enhancement training scheme, where a maximal information-preserving consistency mechanism was adopted to prevent lesion and structure over-tampering. (2) An RE module was introduced to refine the output of the OT module, improving the flexibility, robustness, and applicability of the system. Our study was the first of its kind to bridge the gap between OT-guided generative models and model-based enhancement frameworks. It is a general approach, adaptable to other structure-preserving medical image enhancement research.  (3) Our extensive experimental results on three large retinal imaging cohorts demonstrated the superiority of our proposed method over unsupervised and state-of-the-art (SOTA) supervised methods.

\begin{figure*}[!t]
    \centering
    \includegraphics[width=\textwidth]{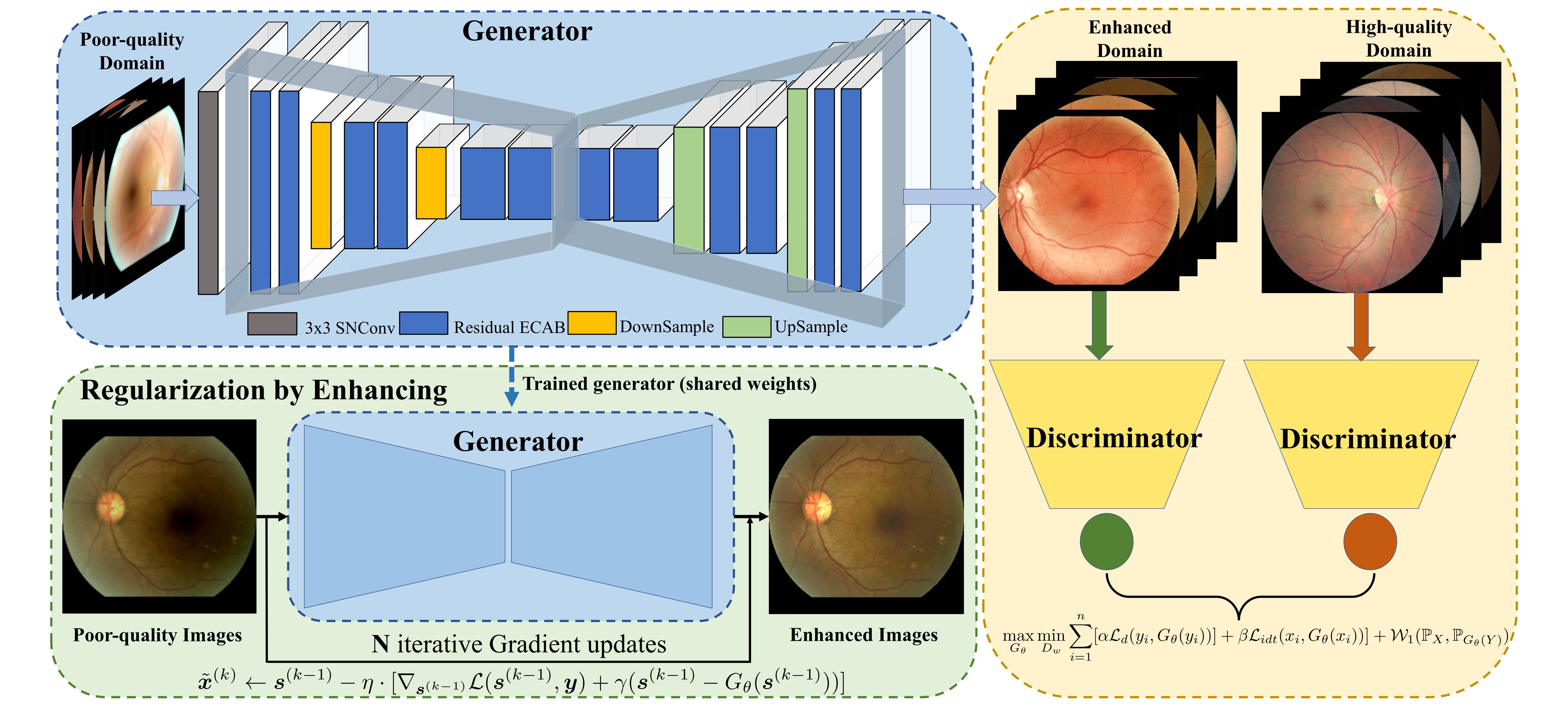}
    \caption{The framework of our proposed method includes OT-guided GAN-based fundus enhancing network and regularization by enhancing framework where SNConv denotes the spectral normalized convolutional layer, and Residual ECAB denotes the residual block with efficient channel attention~\cite{wang2020eca}. The architecture of the discriminator is adopted from~\cite{disc}.}
    \vspace{-1.5em}
    \label{fig:network}
\end{figure*}

\section{Methods}
\vspace{-0.5em}
Restoring clean images $x \sim X$ from their corruptions $y \sim Y$ can be formulated as a variational regularization in the Bayesian framework
\begin{equation}\label{eqn:problemstatement}
    \begin{split}
        \hat{x} = \argmin \limits_{x} f(x)+R(x),
    \end{split}
\end{equation}
where $f$ is the data fidelity measuring the consistency between the restoration and the corrupted data and $R$ is the regularization/prior term. The modern deep learning-based image restoration seeks to train an end-to-end regressor by minimizing the empirical risk
$    \mathbb{E}_{x, y}[\mathcal{L}(f_{\theta}(y), x)]$,
where $f_{\theta}$ is a neural network parameterized by $\theta$, and $\mathcal{L}$ is the loss function. Recent advances~\cite{pnp,red,8253590} show that plugging a learned image regressor into the model-based restoration framework boosts its performance. 
This work unified the model-based regularization and the learned image restoration regressor to provide a flexible, robust framework for enhancing low-quality retinal fundus images. Our framework included two main modules as shown in Fig.~\ref{fig:network}: 1) an OT-guided unsupervised GAN learning scheme serving as a regressor to enhance low-quality images to pursue $f_\theta$ in Eqn.~\ref{eqn:problemstatement}, and 2) an explicit regularization term RE as $R(x)$,  refining the trained generator networks obtained in the first module. The two modules were cascaded together. The entire framework iterated until both modules converged. 

\vspace{-0.5em}
\subsection{OT-Guided Unpaired Image-to-Image Translation}
Let $\mu \sim \mathbb{P}_X$ and $\nu \sim \mathbb{P}_Y$ be two probability measures on the target and source probability manifolds, respectively. The \emph{Monge's} optimal  transport problem of transporting masses from domain $Y$ to $X$ ($Y \rightarrow X$) can be defined as 
\begin{equation}\label{eqn:monge_domain}
    \begin{split}
        \inf \int_{Y} C(y, T(y)) d v(y)
    \end{split}
\end{equation}
where $C(\cdot , \cdot)$ is the cost of transporting $y$ to $T(y)$. The minimal cost among all possible $\nu$-mensurable mappings $T$ yields the optimal transport $u = T^{*}(\nu)$. Intuitively, the transport defined in Eqn.~\ref{eqn:monge_domain} matches the objective of Image-to-Image translation which seeks an optimal mapping from the source domain to the target domain, which we define as \emph{Domain} transport. 
We turn the proposed OT-guided Image-to-Image translation scheme into an optimization problem.
\begin{definition}
The Image-to-Image translation from a source to a target domain $Y \rightarrow X$ suggested by the optimal mass transport can be expressed as 
\begin{equation}\label{eqn:ot_im2im}
    \inf \int_{Y} C(y, T(y)) d v(y), \ \
     \textbf{subject to} \ \ u = T^{*}(\nu)
\end{equation}
\end{definition}
By further parameterizing the optimal transport map $T$ as a neural network $T_{\theta}$, the Eqn.~\ref{eqn:ot_im2im} can be discretized as 
\begin{equation}\label{eqn:opt_im2im}
    \min \limits_{\theta} \mathbb{E}_{y \sim \mathbb{P}_Y}[C(y, T_{\theta}(y))], \ \
    \textbf{subject to} \ \ \mathbb{P}_{T_{\theta}(Y)} = \mathbb{P}_{X}
\end{equation}
By applying the \emph{Lagrange Multiplier}, Eqn.~\ref{eqn:opt_im2im} is relaxed to a constrained optimization given by
\begin{equation}\label{eqn:final_im2im}
    \begin{split}
        \min \limits_{\theta} \mathbb{E}_{y \sim \mathbb{P}_Y}[&C(y, T_{\theta}(y))] + \lambda d(\mathbb{P}_{T_{\theta}(Y)}, \mathbb{P}_{X}),
    \end{split}
\end{equation} 
Likewise, transporting a given measurement in the target domain will also produce another measurement in the target domain $X$, However, we do not desire discrepancies between the measurements on the target. An \emph{Identity} cost constraint is introduced to prevent the network from over-learning or generating unexpected measurements. Meanwhile, it is utilized for maintaining consistency in the target domain. Adding this term to Eqn.~\ref{eqn:final_im2im} can be expressed as:
\begin{equation}\label{eqn:final_6}
    \begin{split}
        \min \limits_{\theta} \mathbb{E}_{y \sim \mathbb{P}_Y}[&C(y, T_{\theta}(y))] + \mathbb{E}_{x \sim \mathbb{P}_X}[C(x, T_{\theta}(x))] + \lambda d(\mathbb{P}_{T_{\theta}(Y)}, \mathbb{P}_{X}),
    \end{split}
\end{equation}
which is defined as \emph{Identity} constrain. where $d(\cdot , \cdot)$ measures the divergence of $\mathbb{P}_{X}$ and $\mathbb{P}_{T_{\theta}(Y)}$, and $\lambda$ is a weight parameter. It is noteworthy that we use the same cost representation, but the \emph{Identity} term is utilized as a constraint in the target domain $X$ and is not related to the optimal transport between the source and target domain.
\vspace{-0.5em}
\begin{proposition}
Supposing Wasserstein-1 distance $\mathcal{W}_1(\cdot, \cdot)$ is applied to measure the divergence between $\mathbb{P}_{X}$ and $\mathbb{P}_{T_{\theta}(Y)}$, Eqn.~\ref{eqn:final_6} suggests an adversarial training scheme of unpaired Image-to-Image translation from $Y \rightarrow X$, given by  
\begin{equation}\label{eqn:adv_im2im}
    \begin{split}
     \max \limits_{G_{\theta}} &\min \limits_{D_{w}} \mathbb{E}_{Y}[\mathcal{L}_d(y, G_{\theta}(y))] + \mathbb{E}_{X}[\mathcal{L}_{idt}(x, G_{\theta}(x))] + \lambda \mathcal{W}_1 (\mathbb{P}_{X}, \mathbb{P}_{G_{\theta}(Y)}) \\
    &  \mathcal{W}_1(\mathbb{P}_{X}, \mathbb{P}_{G_{\theta}(Y)}) = \sup \limits_{||D_{w}||_L \leq 1} \mathbb{E}_{X} [D_{w}(x)] - \mathbb{E}_{Y} [D_{w}(G_{\theta}(y))]
    \end{split}
\end{equation}

where $G_{\theta}$ is the generator parameterized by $\theta$, $D_w$, the discriminator, is a 1-Lipschitz function parameterized by $w$, and $\mathcal{L}_d$ and $\mathcal{L}_{idt}$ denotes the domain transport cost and identity constraint cost, respectively. To better preserve the important structure (e.g., lesions), we ensured that the unpaired input with the matched disease labels $g_i$ if there were any, where $g$ denotes the disease type.
\end{proposition}
The \emph{1-Lipshcitz} constraint is approached by the gradient penalty~\cite{gp} in our experiment. The \emph{Domain} transport and the \emph{Identity} constraint shares the same cost function, as detailed below. 

\begin{algorithm}[t]
\caption{OT-Guided Unpaired Image-to-Image Translation.}\label{alg:ot}
\begin{algorithmic}
\Require The learning rate $\eta$, the batch size $m$, the gradient penalty weight $\lambda$, the consistency loss weight $\alpha \leq 1$, the identity loss weight $\beta$.
\Require Initial discriminator parameters $w_0$, initial generator parameters $\theta_0$.
\While{not converge}
\State Sample a batch of low-quality images $\bm{y} = \{y_{i}\}_{i=1}^{m} \sim \mathbb{P}_{Y}$ with $\{ g_i \}_{i=1}^{m}$.
\State Sample a batch of high-quality images $\bm{x} = \{x_{i}\}_{i=1}^{m} \sim \mathbb{P}_{X}$ with $\{ g_i \}_{i=1}^{m}$.
\For{$i= 1, \dots, m$}
\State Sample a random $\epsilon \sim U[0,1 ]$.
\State $\tilde{x}_{i} \leftarrow G_{\theta}(y_{i})$
\State $\hat{x}_{i} \leftarrow \epsilon x_{i} + (1 - \epsilon) \tilde{x}_{i}$
\State $\mathcal{L}_{D_w} (i) \leftarrow D_{w} (\tilde{x}_{i}) - D_{w} (x_{i})  + \lambda (||\nabla_{\hat{x}_{i}} D_{w} (\hat{x}_{i})  ||_2-1)_{+}^2 $
\EndFor
\State $w \leftarrow w + \eta \cdot \text{RMSProp}(w, \nabla_w \frac{1}{m} \sum_{i=1}^{m} \mathcal{L}_{D_w }(i))$ 
\State $\mathcal{L}_{G_{\theta}} \leftarrow \frac{1}{m} \sum_{i=1}^{m} -D_w(G_{\theta}(\bm{y})) + \alpha \mathcal{L}_d(\bm{y}, G_{\theta}(\bm{y})) + \beta \mathcal{L}_{idt}(\bm{x},  G_{\theta}(\bm{x})) $ 
\State $\theta \leftarrow \theta - \eta \cdot \text{RMSProp}(\theta, \nabla_{\theta} \mathcal{L}_{G_{\theta}} ) $
\EndWhile
\end{algorithmic}
\end{algorithm}

\vspace{-1em}
\subsubsection{Information-Preserving Consistency Mechanism}
There are two main concerns of the proposed OT-guided unpaired image-to-image translation in our task: 1) maintaining the underlying information, e.g., optical discs, lesions, and vessels, consistency before and after the translation;  2) minimizing the duality gap between the primal problem (Eqn.~\ref{eqn:opt_im2im}) and the dual problem (Eqn.~\ref{eqn:final_im2im}). We will introduce our information-preserving consistency mechanism centered on addressing those two main concerns. 

CycleGAN addresses the first concern by introducing the $L_1$ norm as the loss function to enforce low-frequency consistency leading us to the optimal median. In addition, a Patch Discriminator is incorporated to capture high-frequency components by enforcing local structural consistency at a patch level. The Patch Discriminator needs to specify architecture with a pre-defined receptive field usually resulting in a ``shallow" discriminator. 
Our early experiments with CycleGAN, however, indicated that it destroyed the lesion structures and introduced non-existing vessels. Inspired by the Patch Discriminator, we used the multi-scale structural similarity index measure (SSIM)~\cite{pmid22042163} as our consistency loss function $\mathcal{L}_d$ given by 
   $ \mathcal{L}_d(y, G_{\theta}(y)) = 1- \mathbf{SSIM}_{MS}(y, G_{\theta}(y)).$
Followed by the CycleGAN, we also incorporated the identity loss $\mathcal{L}_{idt}$ to make sure that a high-quality input would result in a high-quality enhancement given by
    $ \mathcal{L}_{idt}(x, G_{\theta}(x)) = 1- \mathbf{SSIM}_{MS}(x, G_{\theta}(x)).$
We also used the UNet~\cite{unet} as our generator to help the low-level semantics flow from the poor-quality domain to the high-quality domain. The following theorem~\cite{pmid22042163} provides a theoretical guarantee to our loss function definition. 

\begin{theorem}[\cite{pmid22042163}]
The Structural Similarity Index Measure is proven to be locally Quasi-Convex which minimizes the duality gap between the primal and dual problem and weak duality holds. 
\end{theorem}

To better balance identity loss, domain loss, and the divergence between $\mathbb{P}_X$ and $\mathbb{P}_{G_{\theta}(y)}$, the final objective function was rewritten as 
\begin{equation}
    \vspace{-0.5em}
    \max \limits_{G_{\theta}} \min \limits_{D_{w}}  \sum_{i=1}^{n}[\alpha\mathcal{L}_d(y_i, G_{\theta}(y_i)) + \beta \mathcal{L}_{idt}(x_i, G_{\theta}(x_i))] + \mathcal{W}_1 (\mathbb{P}_{X}, \mathbb{P}_{G_{\theta}(Y)}),
\end{equation}
where $\alpha, \beta$ are weight parameters of the domain loss and identity loss, respectively. The algorithm of our OT-guided unpaired image-enhancing training scheme is given by Algorithm~\ref{alg:ot}.
\vspace{-0.5em}
\subsection{Regularization by Enhancing}
\vspace{-0.5em}
Regularization by Denoising(\emph{RED})~\cite{red} is an off-the-shelf model-based framework that can take advantage of a variety of existing CNN priors without modifying the model's architecture to guide image restoration. We generalized the denoiser-centered RED idea to a more generic one that leveraged our image prior learned from the proposed OT-guided enhancing networks. We formulated the enhancement as an image prior to guiding the restoration of any test images whenever there are not enough samples for the end-to-end training. The objective of our proposed regularization by Enhancing (\emph{RE}) is given by
\begin{equation}
    \begin{split}
        \hat{x} = \argmin \limits_{x} \mathbb{E}_x[\mathcal{L}(x, y)] + \gamma R(x) \ \text{with} \ R(x) = \frac{1}{2}x^T (x - G_{\theta}(x)),
    \end{split}
\end{equation}
where $\gamma$ controls the regularization strength, and $\mathcal{L}$ denotes the multi-scale structural similarity loss. The gradient of the \emph{RE} prior has  a simple form 
\begin{equation}
    \begin{split}
        \nabla_x R(x) = x - G_{\theta}(x),
    \end{split}
\end{equation}
under the condition that $G_{\theta}$ is locally homogeneous and has a symmetric Jacobian. The 1-Lipschitz constraint of $G_{\theta}$ can further guarantee the passivity of $G_{\theta}$ resulting in a convex objective function. We regularized the spectral radius of the weight of each convolutional layer in our generator $G_{\theta}$ via spectral normalization~\cite{sn} to approximate the 1-Lipschitz constraint. In the optimization phase, the accelerated gradient descent was chosen to iteratively approach the optimum. The iterative optimization of the \emph{RE} is given by Algorithm~\ref{alg:re}.

\begin{algorithm}[t]
\caption{Regularization by Enhancing.}\label{alg:re}
\begin{algorithmic}
\Require The step size $\eta$, regularization strength $\gamma$, tolerance \textbf{tol}, Generator $G_{\theta}$
\Require Initial $\tilde{\bm{x}}^{(0)}, \bm{s}^{(0)} = \tilde{\bm{x}}^{(0)}$, $t^{(0)}=1$
\While{not converge}
\State $t^{(k)} = \frac{1}{2} (1 + \sqrt{1 +  4 (t^{(k-1)})^2})$
\State $\textbf{Der}(\bm{s}^{(k-1)}) = \nabla_{\bm{s}^{(k-1)}} \mathcal{L}(\bm{s}^{(k-1)}, \bm{y}) + \gamma (\bm{s}^{(k-1)} - G_{\theta}(\bm{s}^{(k-1)})) $  
\State $\tilde{\bm{x}}^{(k)} \leftarrow \bm{s}^{(k-1)} - \eta \cdot \textbf{Der}(\bm{s}^{(k-1)})$ 
\State $\bm{s}^{(k)} \leftarrow \tilde{\bm{x}}^{(k)} + \frac{t^{(k-1)-1}}{t^{(k)}}(\tilde{\bm{x}}^{(k)} - \tilde{\bm{x}}^{(k-1)}) $
\If{$|| \tilde{\bm{x}}^{(k)} - \tilde{\bm{x}}^{(k-1)} || \leq \textbf{tol} \cdot ||\tilde{\bm{x}}^{(k-1)}||$} 
    \State \emph{break}
\EndIf 
\EndWhile
\end{algorithmic}
\end{algorithm}

\vspace{-0.5em}
\section{Experimental Results}
\vspace{-0.5em}
We conducted extensive experiments in scenarios where the ground-truth clean images are available (\emph{full-reference assessment}) and unavailable (\emph{no-reference assessment}). Three downstream tasks  including DR grading, vessel segmentation, and lesion segmentation, were studied to further evaluate the performance of our proposed method. Visual inspection was conducted by human ophthalmologists to evaluate the performance of no-reference assessment. The vanilla ResNet-50~\cite{He_2016_CVPR} and UNet~\cite{unet} were used to train and test the downstream tasks.

\vspace{-0.5em}
\subsection{Datasets}
 Our proposed method was extensively evaluated on three publicly available retinal CFP datasets: the EyeQ dataset~\cite{eyeQ}, the DRIVE dataset~\cite{drive}, and the IDRID dataset~\cite{idrid}. The EyeQ dataset was manually labeled into three quality levels: good, usable, and reject. We used 7886 training images and 8161 testing images (good \& reject) in our training and evaluation. The DRIVE dataset evaluated our proposed method on the vessel segmentation task with 40 subjects. The IDRID dataset containing 81 subjects with pixel-level annotation of microaneurysms (MA), soft exudates (SE), hemorrhages (HE), and hard exudates (EX) were used to evaluate our method on DR lesion segmentation. All images are center-cropped and resized to a size of $256 \times  256$.
\vspace{-0.5em}
\subsection{Experimental Design}
For the no-reference assessment, our proposed OT-guided Image-to-Image translation GAN  was trained with 7886 training images on the EyeQ dataset by unpaired low-quality images and high-quality images. It was defined as low-quality to high-quality (\emph{low2high}) model. For the full-reference assessment, the model was trained on the subset of the high-quality EyeQ training dataset with degraded images obtained by~\cite{shen2020modeling} and unpaired high-quality images. It was defined as degradation to high-quality (\emph{deg2high}) model. The disease label $g$ in Algorithm~\ref{alg:ot} was the DR grading label from the EyeQ.
Data augmentation, including random horizontal/vertical flips, random crops, and random rotations, was performed to prevent over-fitting during training. All models were trained with the RMSprop optimizer for 200 epochs with an initial learning rate of $1\times$ $10^{-4}$ for the discriminator and $5\times10^{-5}$ for the generator with a decay of 10 by every 100 epochs. The optimal hyperparameters were $\alpha=60$, $\beta=20$  for both low2high and deg2high models. In the testing phase, the optimal hyperparameter $\gamma$ was grid-searched within a range from $1\times10^{-3}$ to $1\times10^{-4}$  with the number of iterations equal to 400 for all experiments. All methods were implemented in PyTorch, and the code will be available on GitHub after the paper acceptance.

\begin{figure}[!t]
\begin{minipage}[b]{1.0\linewidth}
  \centering
  \centerline{\includegraphics[width=\textwidth]{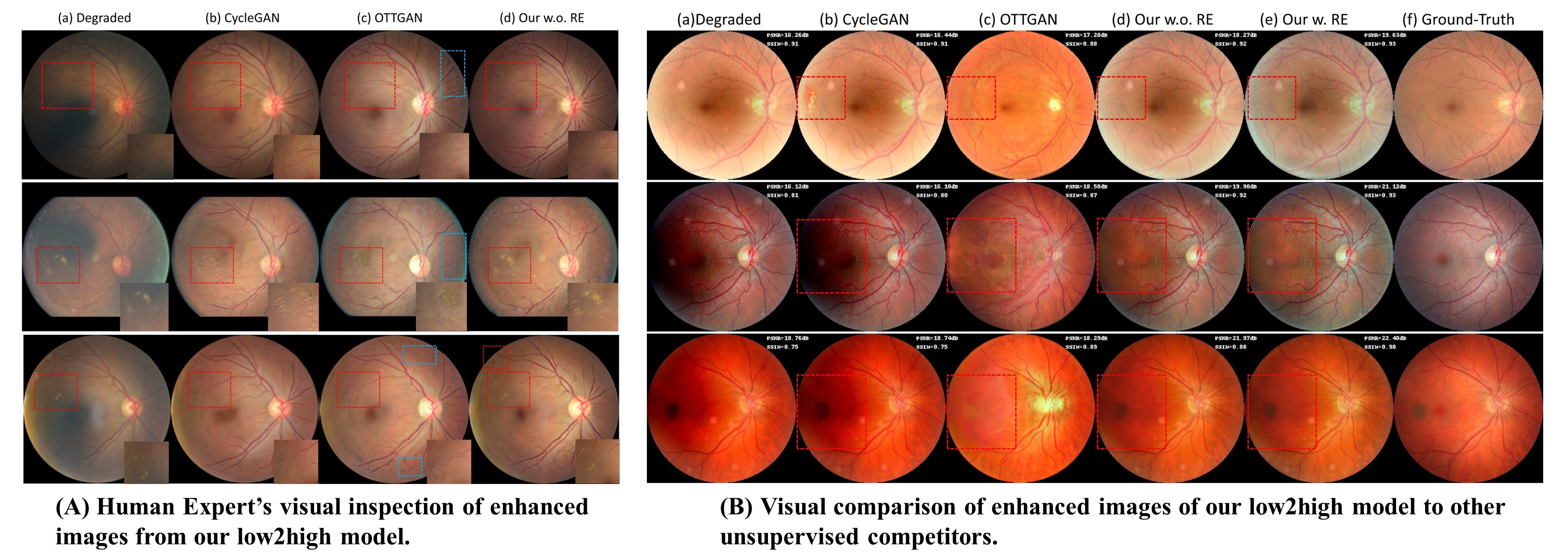}}
  \vspace{-0.5em}
\end{minipage}
\caption{(A).The highlight blocks denote that contrast the structure of the lesion and vessel. It is very obvious that all other methods changed the structure of the lesion or vessel. (B). The red blocks denote generate extra structure compared with our method and areas where noise reduction is not obvious.}
\label{fig:combo1}
    \vspace{-0.5cm}
\end{figure}

\vspace{-0.5em}
\subsection{No-Reference Quality Assessment}\label{sec:low2high}
\begin{table}[!t]
    \centering
    \begin{tabular}{p{1.8cm}P{1.5cm}  P{1.5cm}  P{1.5cm}|P{1.5cm}P{1.5cm}P{1.5cm}}
        \toprule
            \multicolumn{1}{c}{}  & \multicolumn{3}{c}{DR Grading} & \multicolumn{3}{c}{Experts Evaluation}\\
             \cmidrule(lr){2-4}  \cmidrule(lr){5-7}   
         Method & Accuracy &  Kappa &  ROC & LCR & BCR & GESR \\
        \midrule
        CycleGAN & 0.7148 & 0.5378 & 0.9083 & 0.449 & \underline{\textbf{0.0}} & 0.347  \\
        OTTGAN & 0.6996 & 0.5105 & 0.8995 & 0.429 & 0.102 & 0.490\\
        OTRE &  \underline{\textbf{0.7767}} &   \underline{\textbf{0.6814}} &  \underline{\textbf{0.9403}} & \underline{\textbf{0.020}} &  0.040 & \underline{\textbf{0.326}} \\
        \bottomrule
    \end{tabular}
    \caption{Evaluation metrics of the DR grading task with enhancements from different methods in the Lesion structure changed ratio (LCR), background-color changed ratio (BCR), and generated extra structures ratio (GESR).}
    \vspace{-0.9cm}
    \label{tb:DR}
\end{table}
Evaluating the quality of the enhancement without knowing the ground-truth clean images is challenging. We considered combining the DR grading task with visual inspection by human experts to assess the performance of the enhancement. The DR grading task can be viewed as a criterion to judge whether lesion information is preserved after the enhancement. A ResNet-50 model was trained on high-quality images following the experimental setup in~\cite{sipaim} and evaluated by the low-quality images and their enhancements from different methods. The performance of the enhancement will be indicated by the classification accuracy, Area under Receiver Operating Characteristic Curve (AU-ROC), and Cohen's Kappa Coefficient (kappa). 

For visual inspection by human experts, 50 low-quality images were randomly chosen and processed by different enhancement methods. Visual inspection was done to measure 1) the ratio of changing lesion structure (LCR), 2) the ratio of changing the main background color (BCR), and 3) the ratio of generating non-existing structures (GESR). For the fairness of our experiments, the assessment was first conducted by three volunteers who were pretrained with the designed protocol and then finalized by the ophthalmologist.

\begin{table}[!b]
\vspace{-0.5cm}
    \centering
    \begin{tabular}{p{2.5cm}p{3.5cm}cccccc}
        \toprule
         \multirow{2}[3]{*}{} & \multirow{2}[3]{*}{}  & \multicolumn{2}{c}{EyeQ} & \multicolumn{2}{c}{DRIVE} & \multicolumn{2}{c}{IDRID}  
            \\ \cmidrule(lr){3-4}  \cmidrule(lr){5-6} \cmidrule(lr){7-8}
          
        & Method & PSNR &  SSIM & PSNR &  SSIM & PSNR &  SSIM \\
        \midrule
        Supervised &cofe-Net & 23.11 & \underline{\textbf{0.910}} & 21.87 & 0.767 & 20.25 & 0.825 \\
        \cline{1-8}

        
        &CycleGAN \break \textbf{(low2high)} & 18.57 & 0.836 & 18.78 & 0.705 &19.13 & 0.799\\
        &CycleGAN \break \textbf{(deg2high)} & 22.75 & 0.895 & 21.92 & 0.766 & 21.56 &   0.855\\
        
        & OTTGAN \break \textbf{(low2high)} & 18.93 & 0.859 & 19.20 & 0.723 & 19.70 & 0.828 \\
        
        & OTTGAN \break \textbf{(deg2high)} & 23.69 & 0.894 & 21.61 & 0.750 & 21.93 & 0.839 \\
         \cline{2-8}
         Unsupervised& OTRE without RE  & 20.39 & 0.878 & 20.03 & 0.733 & 20.50 & 0.837 \\
        & \textbf{(low2high)} & & & & & & \\
        & OTRE with RE  & 21.08 & 0.880 & 20.61 & 0.740 & 20.55 & 0.836 \\
        & \textbf{(low2high)} & & & & & & \\
        & OTRE without RE  & 24.29 & 0.906 & 22.40 & 0.772 & 21.51 & \underline{\textbf{0.860}}\\
        & \textbf{(deg2high)} & & & & & & \\
        & OTRE with RE  & \underline{\textbf{24.63}} & 0.905 & \underline{\textbf{22.81}} & 
        \underline{\textbf{0.794}}  & \underline{\textbf{22.05}} & 0.852 \\
        & \textbf{(deg2high)} & & & & & & \\
        \bottomrule
    \end{tabular}
    \caption{Result comparison of unsupervised methods when trained with the no-reference training data (Sec.~\ref{sec:low2high}, low2high) and full-reference training data (deg2high) on the current degrading testing dataset. The OTRE frameworks with/without RE module were investigated on both datasets. The supervised method (coef-Net) was trained/evaluated with the degrading dataset only.}
    \label{tb:FULL}
\end{table}

Fig.~\ref{fig:combo1} (A) illustrates some results from different unsupervised enhancement methods. All methods can enhance image quality. However, our method can better maintain the lesion and vessel structure while reducing the noise. We introduced two experiments to verify that our method can preserve the maximal information (Table~\ref{tb:DR}). First we applied DR grading algorithm to the enhanced images (low2high) and evaluated their grading accuracy. As shown in Table~\ref{tb:DR}, the OTRE outperformed other methods in all three measures, expecially by more than $20\%$ in the kappa measure. The human expert inspection also verified that our method could maximize information preservation. Our method performed best in LCR and GESR, with a dramatic improvement in LCR, even over $40\%$ improvement over the other two methods. 

\begin{figure}[!t]
\begin{minipage}[b]{1.0\linewidth}
  \centering
  \centerline{\includegraphics[width=\textwidth]{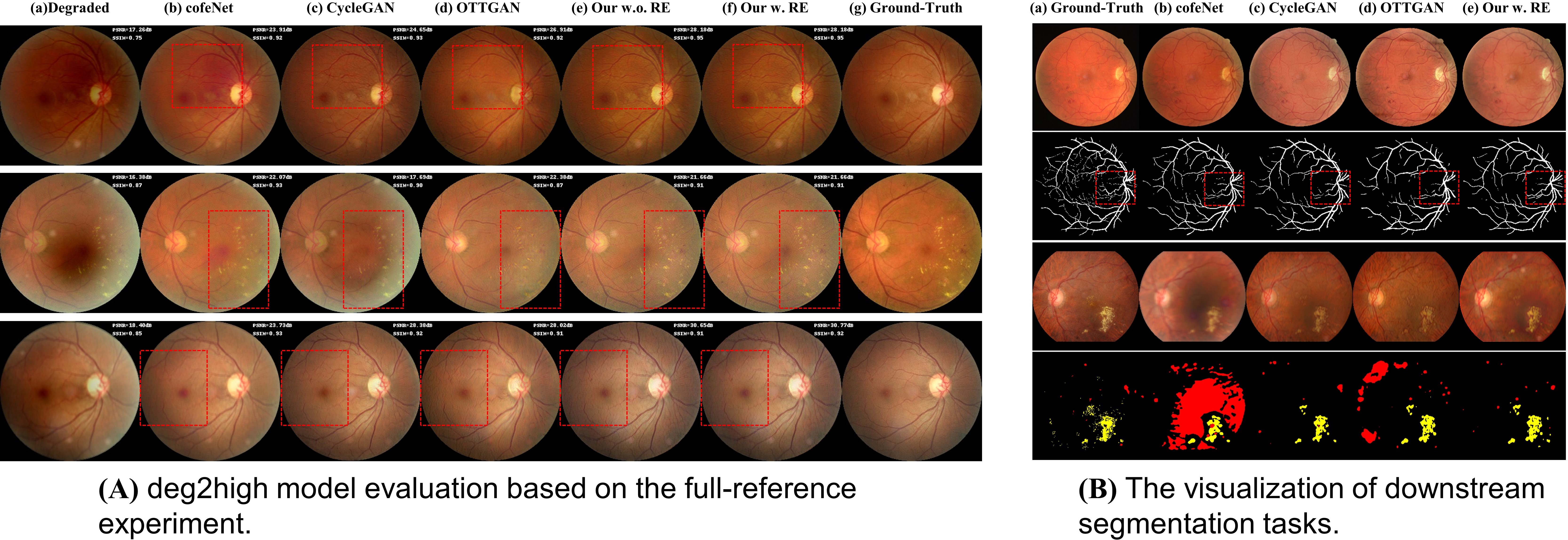}}
  \vspace{-0.5em}
\end{minipage}
\caption{(A). The highlighted red block shows the comparison of structure-preserving and extra structure generation. Visually the image enhancement is good, but the PSNR is not high, so further downstream task evaluation is essential. (B). The highlighted red blocks denote the comparison of fine vessel bifurcation segmentation results.}
\vspace{-0.4cm}
\label{fig:combo2}
\end{figure}

\begin{table}[!t]
    \centering
    \begin{tabular}{p{1.4cm}ccccc|ccc|ccc}
        \toprule
         \multirow{2}[3]{*}{} & \multicolumn{5}{c}{Vessel Segmentation} & \multicolumn{3}{c}{EX} & \multicolumn{3}{c}{HE}
            \\ \cmidrule(lr){2-6}  \cmidrule(lr){7-9}  \cmidrule(lr){10-12} 
          
         Method & ROC &  PR &  F1  &  SE &  SP & ROC &  PR &  F1 &  ROC &  PR &  F1 \\
        \midrule
        cofe-Net & 0.923 & 0.787 & 0.714 & 0.644 &   \underline{\textbf{0.977}} & 0.926 & 0.442 & 0.469 & 0.807 & 0.103 & 0.090 \\
        CycleGAN & 0.910 & 0.762 & 0.696 & 0.622 &  0.975 & 0.900 & 0.474 & 0.347 & 0.845 & 0.155 & 0.141\\
        OTTGAN & 0.900 & 0.739 & 0.667 & 0.581 & 0.976 & 0.912 & 0.507 &  \underline{\textbf{0.512}} & 0.855 & 0.107 & 0.145 \\
        OTRE &   \underline{\textbf{ 0.927}} &    \underline{\textbf{0.796}} &   \underline{\textbf{0.726}} &  \underline{\textbf{0.672}} & 0.975 & \underline{\textbf{0.934}} & \underline{\textbf{0.529}} & 0.441 & \underline{\textbf{0.894}}& \underline{\textbf{0.233}} & \underline{\textbf{0.273}} \\
        \bottomrule
    \end{tabular}
    \caption{Result comparison of the segmentation of blood vessels on the DRIVE cohort~\cite{drive} and diabetic lesions (EX and HE) on the IDRID dataset~\cite{idrid}.  The OTRE compared favorably to other supervised and unsupervised methods. (ROC: Area under Receiver Operating Characteristic Curve, PR: Area under the Precision-Recall, F1: F1 score, SE: Sensitivity, SP: Specificity).}
        
        \vspace{-1.5em}
    \label{tab-seg}
\end{table}

\vspace{-0.3cm}
\subsection{Full-Reference Quality Assessment}
\vspace{-0.3em}
For full-reference assessment, we degraded high-quality images following the degradation model introduced by Shen at.al~\cite{shen2020modeling} to synthesize the low-quality images for each dataset. The training dataset consisted of 6500 high-quality images selected from the EyeQ training dataset and other 6500 synthesized low-quality images degraded from non-overlapping 6500 high-quality images from the EyeQ training dataset. The testing dataset was made up of 500 images from the EyeQ testing dataset, the entire DRIVE dataset, and the entire IDRID dataset. The commonly used Peak-Signal-to-Noise Ratio (PSNR) and Structural Similarity Index Measure (SSIM) were used to evaluate the quality of the enhanced low-quality images. To further validate our proposed method on the downstream tasks, we evaluated the performance of our proposed method on the blood vessel segmentation and DR lesion segmentation tasks. 

First, we evaluated the consistency of the enhanced images and their high-quality counterparts. We performed two different experiments by applying both no-reference trained models (Sec.~\ref{sec:low2high}, low2high) and full-reference trained models (deg2high). We also tested whether the inclusion of RE module improved our results. Fig.~\ref{fig:combo1} (B) and Fig.~\ref{fig:combo2} (A) show some image examples and Table~\ref{tb:FULL} reports the numerical results. As shown in Table~\ref{tb:FULL}, except for EyeQ's SSIM measure, our OTRE outperformed all other supervised and unsupervised methods in three different datasets, and the PSNR achieved respectively the highest 24.63, 22.81, and 22.05. Remarkably, the OTRE beat the SOTA supervised method (cofe-Net) given that the cofe-Net was trained with paired images, but the OTRE was not. Interestingly, our method no-reference trained model (\emph{low2high}) achieved competitive results for the unseen degradation noises. We also learned that the inclusion of RE module gained improved performance. Fig.~\ref{fig:combo1}(B), Fig.~\ref{fig:combo2}(A) also provided stronger support of effectiveness that our method preserved the structure and achieved better noise reduction. 

To further confirm the superiority of our method, two downstream segmentation tasks were studied using the groundtruth data from DRIVE and IDRID datasets. Since the training and testing of our segmentation task were based entirely on enhanced images, without adding any preprocessing and additional tricks, in the lesion segmentation task, we only considered large blocks of lesions which were easy to train, such as EX and HE. As shown in Table~\ref{tab-seg}, the OTRE method achieved excellent results in three segmentation tasks. It achieved the highest ROC and PR results in all segmentation results, 2 out of 3 bests results in F1 measure. From some image examples shown in Fig.~\ref{fig:combo2} (B), it is easy to see that other methods have the problem of insignificant enhancement performance, resulting in the altered vessel and lesion structures.




\vspace{-0.5cm}
\section{Conclusion and Future Work}
\vspace{-0.3em}

This work integrated OT-guided GAN-based enhancing network with the RE module and achieved promising results on three datasets, surpassing or on a par with SOTA unsupervised and supervised methods. A limitation of the current system is that it assumed all input data were usable but in real clinical applications, there exist some images that are completely corrupted. A screening procedure to classify whether the input images are usable will make our work more practical. We will study it in our future work.


\noindent\paragraph{\textbf{Acknowledgement.}} This work was partially supported by grants from NIH (R21AG065942, R01EY032125, and R01DE030286).

\bibliographystyle{splncs04}
\bibliography{refs}

\end{document}